%% file: main_arxiv.tex
\documentclass{article}

\usepackage{arxiv}

\usepackage[T1]{fontenc}    
\usepackage{newtxtext,newtxmath}
\usepackage{hyperref}       
\usepackage{url}            
\usepackage{booktabs}       
\usepackage{alltt}
\usepackage{float}
\usepackage{amsmath}
\usepackage{amsfonts}       
\usepackage{nicefrac}       
\usepackage{microtype}      
\usepackage{graphicx}
\usepackage[authoryear]{natbib}
\usepackage{caption}
\usepackage{subcaption}
\usepackage{bm}
\usepackage{listings}

\usepackage{algorithm}
\usepackage{algpseudocode}
\usepackage{enumitem} 

\newcommand\independent{\protect\mathpalette{\protect\independenT}{\perp}}
\def\independenT#1#2{\mathrel{\rlap{$#1#2$}\mkern2mu{#1#2}}}

\newcommand{\ra}[1]{\renewcommand{\arraystretch}{#1}}

\usepackage{tikz}
\usetikzlibrary{arrows,automata}
\tikzset{
	semi/.style={
		semicircle,
		draw,
		minimum size=2em
	}
}
\usetikzlibrary{shapes, decorations, graphs, shapes.geometric, positioning,swigs}

\newcommand{\Ebb}{\mathbb{E}}

\newcommand{\e}{\mathrm{e}}
\newcommand{\diff}{\mathop{}\!\mathrm{d}}

\title{revealing the truth: calculating true values in causal inference simulation studies via gaussian quadrature}

\author{ \href{https://orcid.org/0000-0003-1967-039X}{\includegraphics[scale=0.06]{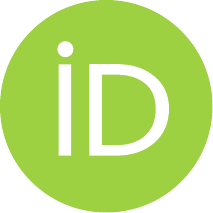}\hspace{1mm}Alex Ocampo}\\
	Statistical Methodology\\
	F.\ Hoffman La Roche\\
    Basel, Switzerland \\
	\texttt{alex.ocampo@roche.com} \\
	\And
	\href{https://orcid.org/0000-?}{\includegraphics[scale=0.06]{orcid.pdf}\hspace{1mm}Enrico Giudice} \\
	Advanced Quantitative Sciences\\
	Novartis Pharma AG\\
    Basel, Switzerland \\
	\texttt{enrico.giudice@novartis.com} \\
 \And
	\href{https://orcid.org/0000-0002-2006-9828}{\includegraphics[scale=0.06]{orcid.pdf}\hspace{1mm}Zachary McCaw} \\
	Department of Biostatistics \\
	University of North Carolina at Chapel Hill \\
    Chapel Hill, NC, USA \\
	\texttt{zmccaw@alumni.harvard.edu} \\
 \And
	\href{https://orcid.org/0000-0001-5850-3610}{\includegraphics[scale=0.06]{orcid.pdf}\hspace{1mm}Tim P.\ Morris} \\
	Statistical Methodology \\
    Novartis Pharma.\ UK Ltd. \\
    London, UK \\
	\texttt{tim.morris@novartis.com} \\
}

\renewcommand{\shorttitle}{Revealing the truth with Gaussian quadrature}

\hypersetup{
    pdftitle={Revealing the truth with Gaussian quadrature},
    pdfsubject={Revealing the truth with Gaussian quadrature},
    pdfauthor={Alex Ocampo, Enrico Giudice, Zachary McCaw, Tim P Morris},
    pdfkeywords={Simulation studies, Causal inference, Gaussian quadrature, Monte Carlo integration},
}

\providecommand{\bsize}{}

\begin{document}

\maketitle 

\input{abstract}

\input{manuscript}

\bibliographystyle{apalike}
\bibliography{bib}

\input{appendices}

\end{document}

%% file: abstract.tex
\begin{abstract}
    \noindent Simulation studies are used to understand the properties of statistical methods. A key luxury in many simulation studies is knowledge of the true value (\textit{i.e.} the estimand) being targeted. With this oracle knowledge in-hand, the researcher conducting the simulation study can assess across repeated realizations of the data how well a given method recovers the truth. In causal inference simulation studies, the truth is rarely a simple parameter of the statistical model chosen to generate the data. Instead, the estimand is often an average treatment effect, marginalized over the distribution of confounders and/or mediators. Luckily, these variables are often generated from common distributions such as the normal, uniform, exponential, or gamma. For all these distributions, Gaussian quadratures provide efficient and accurate calculation for integrands with integral kernels that stem from known probability density functions. We demonstrate through four applications how to use Gaussian quadrature to accurately and efficiently compute the true causal estimand. We also compare the pros and cons of Gauss--Hermite quadrature to Monte Carlo integration approaches, which we use as benchmarks. Overall, we demonstrate that the Gaussian quadrature is an accurate tool with negligible computation time, yet is underused for calculating the true causal estimands in simulation studies.
\end{abstract}


%% file: manuscript.tex
\section{Introduction}

Many simulation studies involve generating data from a known distribution and then considering methods’ abilities to recover parameters. The researcher’s oracle-like knowledge of the true value of target parameters is one reason why simulation studies offer greater insights into methods’ operating characteristics than analyzing a real dataset/s. This paper comes from our experience generating data for causal inference simulation studies. Any such analysis must address the question, “what is the truth”; that is, what is the true value of the causal estimand that our methodology is targeting? A fair evaluation of a method in a simulation study requires this knowledge for evaluating certain fundamental performance measures such as bias, mean squared error, and coverage. If the truth is unknown or approximated poorly, then methods that are unbiased may appear to be biased, and so on.

Simulation studies frequently focus on what might be termed \textit{data-generating model parameter recovery}, where the target of inference is a specified parameter of the data-generating model, so its true value is not in doubt. Obtaining true values is often less straightforward in simulation studies where the target of inference is a causal estimand. For instance, if one simulates from a generalized linear regression model with $\beta=5$, and $\beta$ is the estimand of interest, we know that we should assess bias, $\Ebb(\widehat{\beta})-\beta$, using a true value of 5 for $\beta$. In causal inference, interest revolves around functionals of potential outcome distributions, which may not be derivable from parameters of a data-generating model. For instance, proper confounding control is often important, and therefore we must marginalize (or integrate) functionals over the distributions of all confounders in order to compute the true causal effect of interest. For example, suppose we are interested in the average causal effect
\begin{align*}
    \Ebb[Y^{(a=1)}] - \Ebb[Y^{(a=0)}],
\end{align*}
but need to marginalize over the distribution of a confounder -- variable $C\in\mathbb{R}$ -- in order to calculate these expectations for a given data-generating mechanism for $Y^{(a)}$. That is, for both terms above, we need to perform the following\footnote{Note that the conditioning on $A=a$ in \eqref{eq:conda} is redundant here; it is also harmless, so we retain it for familiarity.}:
\begin{align}
    \mathbb{E}[Y^{(a)}] &= \Ebb[\mathbb{E}[Y^{(a)}|C]] \notag\\
    &= \Ebb[\Ebb[Y^{(a)}|A=a,C]] \label{eq:conda}\\
    &= \int_{-\infty}^{\infty} \Ebb[Y^{(a)}|A=a,C] f(c)dc, \quad \forall a \in \{0,1\}.  \label{eq:c1}
\end{align}
The term $\Ebb[Y^{(a)}|A=a,C]$ is is typically specified as one of the structural equations of the simulation model. These structural equations are frequently specified as a parametric regression model. This allows the (potential) outcome $Y^{(a)}$ to be simulated conditional on the treatment set to $a$ and the observed value of the confounder $C$ (e.g. $\mathbb{E}[Y^{(a)}|A=a,C]=g^{-1}(\beta_0+\beta_aa+\beta_CC$). However, integrating $\mathbb{E}[Y^{(a)}|A=a,C]$ over the distribution of $C$ will not have a simple closed-form expression in general, even for familiar confounder distributions of $C$ when the link function $g(\cdot)$ is nonlinear, for example when $g(\cdot)=\text{logit}(\cdot)$.


One solution to this problem is to simulate a very large sample from the data-generating mechanism, including the outcomes, and analyze it using a method known to be unbiased, returning an estimate to be used as the true value. Nine of the 74 simulation studies reviewed in \citet{Morris2019} where the target was an estimand used this simulation-based approach. Its advantages are 1) it is straightforward for statisticians to implement given the data-generating mechanism and 2) because accuracy of the truth is essentially guaranteed by the law of large numbers. However, there are two disadvantages: first, the `true' value is subject to Monte Carlo error due to finite simulated dataset size; second, generation and analysis of an appropriately large dataset carries a computational cost. Although not a “disadvantage”, a condition of using this approach is that, an estimator must be available that is known to be unbiased for the given data-generating mechanism.

More recently, \citet{Naimi2025} proposed Monte Carlo integration as a more accurate yet still practical solution to this problem. This involves again generating a large sample from the data-generating mechanism for some of the variables. However, rather than simulate observed outcomes, we generate the functional of the potential outcomes that defines the estimand. For example, rather than draw a binary outcome to realize values of 0 or 1, we compute its expected value from the data-generating model. We do this under each value of treatment (or more generally cause) of interest. Finally, simple averages are taken that can be transformed if necessary and contrasted across levels of the treatment/cause. The two disadvantages mentioned for outcome simulation remain but the Monte Carlo error associated with the true value is lower (for a given dataset size); put another way, the same Monte Carlo SE can be achieved with a smaller dataset size, making it a more computationally efficient approach. Further, because the approach contrasts potential outcomes, we do not require access to an appropriate unbiased estimator to use it.

One opportunity to avoid generating very large datasets and accepting Monte Carlo error in the true value is to note that, oftentimes, the variables we need to marginalize over in causal inference simulation studies are generated from a common distribution. For example, considering the integral defined in \eqref{eq:c1}, if $C\sim \mathscr{N}(\mu_c,\sigma^2_c)$ this would mean evaluating
\begin{align}
\label{eq:Cint}
    \int_{-\infty}^{\infty} \mathbb{E}[Y|A=a,C] \phi_C(c)dc, 
\end{align}
where 
\begin{align*}
     \phi_C(c)=\frac{1}{\sqrt{2\pi\sigma_c^2}} e^{-\frac{1}{2\sigma_c^2}(c-\mu_c)^2}    
\end{align*}
is the probability density function (pdf) of the normal distribution of $C$. Integrals of this form can be computed instantaneously and accurately \textit{via} Gauss--Hermite quadrature \citep{gauss1814methodus, hermite1864nouveau, heath2002}. 

This paper applies and evaluates the use of Gaussian quadrature for computing true values of interest in simulation studies for causal inference. We begin with an overview of Gaussian quadratures, with particular emphasis on Gauss--Hermite quadrature, since variables in simulation studies are often generated from normal distributions. We then revisit the two examples in \cite{Naimi2025} and add two further examples from our own simulation studies on mediation analyses. We conclude with a discussion to provide recommendations of when a Monte Carlo integration or Gaussian Quadrature approach may be considered. \texttt{R} code is provided to foster application of Gaussian quadrature in causal simulation studies. 



\section{Gauss--Hermite quadrature}
\subsection{Univariate quadrature}
\label{sec:uniquad}
Gauss--Hermite (G–H) quadrature approximates integrals of the following form with a finite sum:
\begin{equation}
\label{eq:GHQ}
    \int_{-\infty}^{\infty}  f(x) e^{-x^2} \, dx \approx \sum_{i=1}^K w_i f(x_i).
\end{equation}
We refer to the $w_i$ as \textit{weights} and the $x_i$ as \textit{nodes}. The nodes are values spread across the domain of $x$, chosen judiciously as the roots of the $K$-th degree Hermite polynomial $H_K(x)$\footnote{There are two types of Hermite polynomials: the probabilist's and physicist's. Gaussian Quadrature uses the physicist's Hermite polynomials defined as $H_K(x)=(-1)^K e^{x^2}\frac{\partial^n}{\partial x^n}e^{-x^2}$.}. The weights are defined as $w_i=(2^{K-1}K!\sqrt{\pi}) / (K^2[H_K(x_i)]^2)$. The length of the sum $K$ defines the number of quadrature points, which can vary from application to application. A surprisingly small number of points is needed for accurate Gauss--Hermite quadrature; \texttt{R} software typically defaults to $K=5$. This is because G–H quadrature with $K$ nodes provides an exact calculation for polynomials of degree $2K-1$ or lower, which is much more accurate than integration using Riemann sums or trapezoid approximations.

Detail-oriented readers may recognize that normally-distributed variables like $C\sim \mathscr{N}(\mu_c,\sigma^2_c)$ from section 1 do not actually fit this form. However, we can perform a simple change-of-variables $x = (c-\mu_C)/\sqrt{2}\sigma_C$ in order to get it into the correct form
\begin{align*}
    \int_{-\infty}^{\infty} f(c) \phi_C(c) dc =& \int_{-\infty}^{\infty} f(c) \frac{1}{\sqrt{2\pi\sigma_c^2}} \e^{-\frac{1}{2\sigma_c^2}(c-\mu_c)^2}  dc \\
    =& \frac{1}{\sqrt{\pi}} \int_{-\infty}^{\infty} f(x\sqrt{2}\sigma_C+\mu_C)  \e^{-x^2}  dc, \\
\end{align*}
which aligns to the necessary integrand to perform Gauss--Hermite quadrature. Therefore, no matter the parameterization of the normal variable we need to integrate over, we can apply the quadrature. This may require rescaling the quadrature points according to $\mu_C$ and $\sigma_C$; a trivial but important task. Univariate Gauss--Hermite quadrature is available through the \texttt{spatstat.random} package in \texttt{R}.





\subsection{Multivariate quadrature}
\label{sec:mutiquad}

Gauss--Hermite Quadrature can be extended to a multivariate integrand when multiple normal distributions need to be considered. For an excellent overview of multivariate Gauss--Hermite quadrature, see \citet{jackel2005note}. 

To build intuition for the multivariate quadrature, consider the bivariate case where two integrals need to be evaluated. For example,
\begin{align*}
    \int_{-\infty}^{\infty} \int_{-\infty}^{\infty}  f(c_1,c_2) \phi_{C_1}(c_1)\phi_{C_2}(c_2) dc_1 dc_2. 
\end{align*}
This may be the case, if for instance, one needs to integrate over two confounding variables. Instead of obtaining quadrature points across the domain of a single variable, in the bivariate case, we must form a grid of quadrature points across the joint domain of $C_1$ and $C_2$.

\begin{figure}
\centering
\small{\hspace{-0.1cm} No decomposition \hspace{2.9cm} Cholesky  \hspace{3.6cm} Spectral} \\
    \includegraphics[width=.32\textwidth]{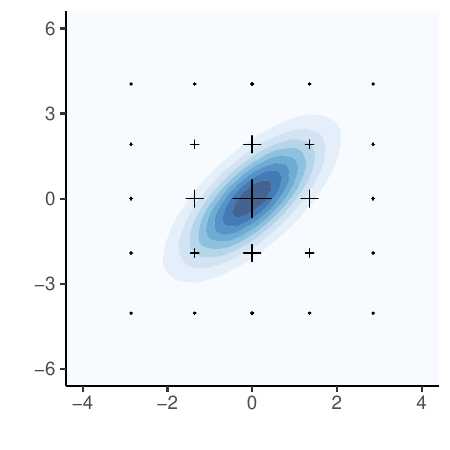}
    \includegraphics[width=.32\textwidth]{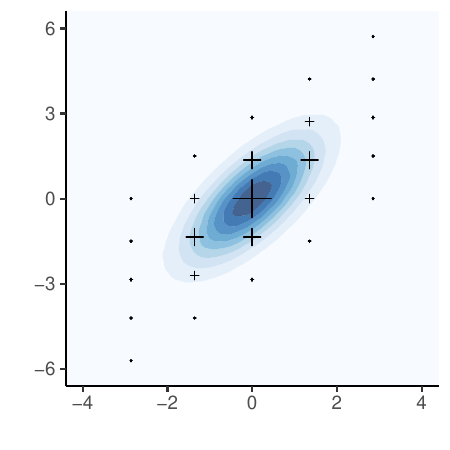}
    \includegraphics[width=.32\textwidth]{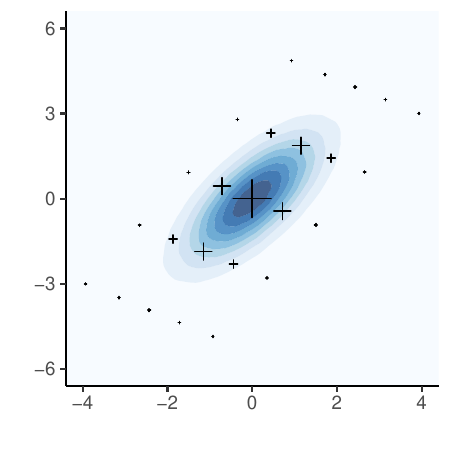}
\caption{\textbf{Grids of quadrature points with $K=5$ levels for a bivariate Gaussian distribution.} Left panel: a simple grid which ignores the correlation between the two variables. Central panel: grid rescaled according to the Cholesky decomposition. Right panel: grid rescaled according to the spectral decomposition of the covariance matrix. The grid points’ sizes are proportional to their corresponding quadrature weights.}
\label{fig:example1grid}
\end{figure}

In standard software, such as the \texttt{R} package \texttt{mvQuad} \citep{citemvquad}, one specifies the level of the quadrature (corresponding to $K$ in the univariate case, e.g. 5, which then leads to $5^2=25$ quadrature points. Therefore, the finite sum that now approximates the double integral is of length 25. Still, even with this increase in quadrature points, multivariate Gauss--Hermite quadrature is still incredibly efficient since computing even large sums is generally computationally simple. In the multivariate space, often na\"ive placement of the points in the quadrature as $\sum_{i=1}^K\sum_{j=1}^K w_i w_j f(c_1,c_2)$ is inefficient when the variables are correlated. To overcome this, one can employ a decomposition which rotates the grid of quadrature points according to the known correlation structure. Incorporating information about the correlation between the confounders optimizes the precision and computational efficiency of the quadrature. \citet{jackel2005note} shows that the Cholesky decomposition is more efficient than the na\"ive grid and the spectral decomposition is more efficient still. These approximate the integrals using 
\begin{align*}
    &\sum_{i=1}^K\sum_{j=1}^K w_i w_j f(c_1,\rho c_1 + \rho^\prime c_2)\\
    \text{and }&\sum_{i=1}^K\sum_{j=1}^K w_i w_j f(ac_1 + bc_2,bc_1 + ac_2) \text{ respectively,}\\
    \text{ where } & \rho = \text{Corr}(c_1,c_2), \quad a = 0.5\left(\sqrt{1+\rho}+\sqrt{1-\rho}\right) \text{ and } b = 0.5\left(\sqrt{1+\rho}-\sqrt{1-\rho}\right).
\end{align*}
Figure~\ref{fig:example1grid} provides a visualizaton of the placement of the grid points for each of these decompositions with $K=5$ and $\rho=\sqrt{0.5}$.

\begin{figure}
\centering
    \includegraphics[width=.3\textwidth]{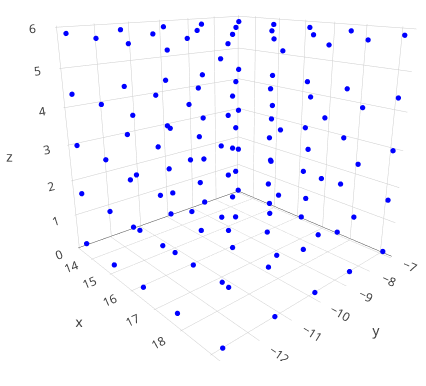} \hspace{1em}
    \includegraphics[width=.3\textwidth]{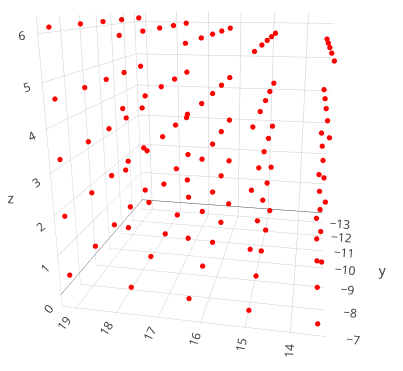}\hspace{1em}
    \includegraphics[width=.3\textwidth]{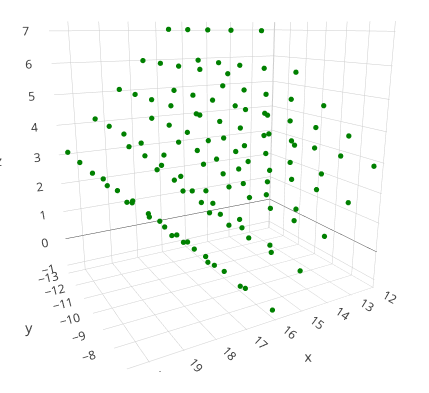}
\caption{\textbf{3-D grids of quadrature points for the a three dimensional multivariate Gaussian distribution.} The simple grid on the left evenly places (blue) quadrature points into a grid forming a square box. The middle figure performs a Cholesky decomposition based on the correlation of the multivariate normal to slightly angle the (red) quadrature points. The right figure rescales the grid according to the spectral decomposition of the covariance matrix which creates a diamond structure. In this final grid, it is clear that only two of the variables are correlated as in one dimension the surface stays constant.}
\label{fig:3D}
\end{figure}

\subsection{Other types of Gaussian quadrature}
\label{sec:othergauss}
So far we have focused on the use of Gauss--Hermite quadrature to approximate integrals of the type (\ref{eq:GHQ}), as continuous variables in simulation scenarios are commonly assigned Gaussian distributions. Nonetheless, there are instances in which non-Gaussian variables might be relevant. Such scenarios may arise when the simulation study investigates distributions with different domains, or when there is interest in exploring how the outcomes are affected by relaxing the Gaussian distribution assumption.

Gauss--Hermite quadrature is a special case of a more general class of methods known as Gaussian quadrature methods. Gaussian quadrature methods accurately approximate specific types of integrals as weighted sums of function values, evaluating the function at specific points. In the context of simulation studies, general Gaussian quadrature methods can be used to compute the expectation of a function $f$ of a continuous random variable with density function $p(x)$: 
\begin{equation*}
    \int f(x)\,p(x)\,dx \,\approx\, \sum_{i=1}^K w_i\,f(x_i).
\end{equation*}
The general Gaussian quadrature problem involves using an orthogonal family of polynomials $G_K(x)$ of degree $K$ such that 
\begin{equation*}
    \int p(x)\,x^j\,G_K(x)\,dx \,=\, 0, \quad\, \forall\, j=0,\dots,n-1.
\end{equation*}
The roots of the polynomial $G_K(x)$ will then be used as quadrature points $x_i$, while the weights $w_i$ are determined by the condition that the quadrature be exact for polynomials up to degree $2K-1$. The weights and evaluation points of the Gaussian quadrature rule are thus uniquely defined by the density $p(x)$: in the case of a density $p(x) \propto e^{-x^2}$, the evaluation points that provide exact results for polynomials $f(x)$ up to order $2K-1$ are given by the roots of Gauss--Hermite polynomials. Table \ref{tab:gaussquads} shows different quadrature rules and the corresponding density kernels with respect to which the function is implicitly being integrated over. For a complete list of polynomials for optimal quadrature associated with different functions $p(x)$, see \cite{Zwillinger02}.

\begin{table}[tbh!]
\centering
\ra{1.4} \addtolength{\tabcolsep}{1.6pt} 
\begin{tabular}{lll} \hline
    {Quadrature rule} & ~{Integral type}~ & {Implicit kernel} \\ \hline
    {Gauss--Hermite} & ~$\int^\infty_{-\infty}\text{e}^{-x^2}f(x)\,dx$~ & {Gaussian}  \\
    {Gauss--Legendre} & ~$\int^1_{-1}f(x)\,dx$~ & {Uniform} \\
    {Gauss--Laguerre} & ~$\int^\infty_0\text{e}^{-x}f(x)\,dx$~ & {Exponential} \\
    {Generalized Gauss--Laguerre} & ~$\int^\infty_0x^\alpha \text{e}^{-x}f(x)\,dx$~ & {Gamma}          \\ \hline
\end{tabular}
\caption{Summary of different quadrature rules, their associated integrals and the corresponding density kernels with respect to which $f(x)$ is being integrated. The generalized Gauss--Laguerre quadrature is defined for any real $\alpha>-1$.}
\addtolength{\tabcolsep}{1pt}
\label{tab:gaussquads}
\end{table}
Details on the above quadrature rules and how these can be used to integrate over generic non-Gaussian densities are deferred to appendix~\ref{app:nongaussian}.

\section{Applications to confounding}

\subsection{Normally distributed confounder}
\label{sec:confex1}
\begin{figure}[htbp]
    \centering
\begin{tikzpicture}
\tikzset{line width=1pt, outer sep=0pt,
         ell/.style={draw,fill=white, inner sep=2pt,
          line width=1pt},
         swig vsplit={gap=5pt,
         inner line width right=0pt}};
\node[name=A, ell, shape=ellipse] {$A$};
\node[name=C, ell, shape=ellipse] at (1,1.5) {$C$};
\node[name=Y, ell, shape=ellipse] at (2,0) {$Y$};
\draw[->,line width=1.5pt,>=stealth,color=black]
(A) edge (Y)
(C) edge (A)
(C) edge (Y);
\end{tikzpicture}    
    \caption{DAG for the simple confounding scenario.}
    \label{fig:dagex1}
\end{figure}
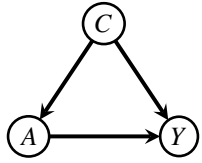

The first example comes from \citet{Naimi2025}, where interest is in the marginal odds ratio, defined as
\begin{align} 
\Psi_{OR} \,=\, \frac{\mathbb{P}(Y^{(1)}=1)}{\mathbb{P}(Y^{(1)}=0)} \,\big/\, \frac{\mathbb{P}(Y^{(0)}=1)}{\mathbb{P}(Y^{(0)}=0)}\,. \label{eq:margor}
\end{align}
To estimate this quantity, we adjust for a univariate, normally-distributed confounder $C$. The data-generating mechanism follows the simple DAG structure in figure~\ref{fig:dagex1}, where the confounder variable follows a normal distribution and the outcome follows a parametric logistic regression model:
\begin{equation} \label{eq:ex1sim}
\begin{split}
    C &\,\sim\, \mathscr{N}(\mu_c, \sigma_c^2) \\
    \mathbb{P}(Y=1\,\vert\,A,C) &\,=\, \textrm{expit}(\beta_0 + \beta_1 A + \beta_2 C). 
\end{split}
\end{equation}

Note that the parametric model for the exposure $A$ will be required when generating a complete simulated dataset. However, since the potential outcomes in equation (\ref{eq:margor}) intervene on the exposure, the model for $A$ is not required for the purposes of computing the true value $\Psi_{OR}$.

Computing causal quantities under this setting requires solving an integral of the type (\ref{eq:GHQ}) in order to compute $\mathbb{E}[Y^{(a)}]=\mathbb{P}(Y^{(a)}=1)$ for $a=\{0,1\}$. In the example the quantities are then used to derive the marginal odds ratio (\ref{eq:margor}).

Due to the difficulty in analytically evaluating the quantities $\mathbb{P}(Y^{(a)}=y)$,
as shown in \cite{Naimi2025}, Monte Carlo integration and numerical integration can be used to estimate them. The methods employed only allowed for a univariate confounder $C$ due to computational and complexity limitations. Here, we extend this example to a multivariate confounding variable $C$ and apply Gauss--Hermite quadrature to compute the true marginal odds ratio.

The problem involves integrating over a single normally-distributed confounder $C$. As a simple example to illustrate the value of Gauss--Hermite quadrature in this setting, we set the simulation parameters $\beta_0 , \beta_1, \sigma^2_c$ in model (\ref{eq:ex1sim}) to $1$, $\mu_c=0$ and $\beta_2=-1$, then apply Gauss--Hermite quadrature with $K=20$ quadrature points. The following code snippet implements the approach in \texttt{R}:

\begin{alltt}
require(mvQuad)
PrY <- function(c,a) 1/(1+exp(c-a-1))
grid <- createNIGrid(dim = 1, type = "GHN", level = 20)
mu1 <- quadrature(PrY, grid, a = 1)
mu0 <- quadrature(PrY, grid, a = 0)
(mu1/(1-mu1)) / (mu0/(1-mu0))  # odds ratio
\end{alltt}
Alternative choices of $\mu_c$ and $\sigma^2_c$ can be implemented via the \texttt{rescale.NIGrid(m = $\mu_c$, C = $\sigma^2_c$)} function.

We compare the true value obtained by Gauss--Hermite quadrature with that estimated by simulating a large sample $N$ of potential outcomes $Y^{(a)}$ from the data-generating mechanism and taking the sample mean as an unbiased estimator of $\mathbb{P}(Y^{(a)}=1)$:
\begin{equation} \label{eq:mcex1a}
\begin{split}
    \widehat{\mathbb{P}}(Y^{(a)}=1) &= \frac{1}{N} \sum_{j=1}^N Y^{(a)}_j , \\
    \text{where } Y^{(a)}_j|C=c_j &\sim \textrm{Bernoulli}(\textrm{expit}(\beta_0 + \beta_1 a + \beta_2 c_j)) \\
    \text{and } c_j & \sim \mathscr{N}(\mu_c,\sigma^2_c).
\end{split}
\end{equation}
We then estimate the truth $\Psi_{OR}$ by plugging in $\widehat{\mathbb{P}}(Y^{(1)}=1)$ and $\widehat{\mathbb{P}}(Y^{(0)}=1)$ to \eqref{eq:margor}. It is worth noting that the procedure used to determine $\Psi_{OR}$ does not require specific knowledge (e.g.~ normality) of the distribution fo $C$, as long as samples can be generated from it.

We additionally estimate the true odds ratio using the Monte Carlo integration of \cite{Naimi2025}, obtained by evaluating the expected value of the potential outcome across a large number of samples from the distribution of $C$:
\begin{equation} \label{eq:mcex1}
\begin{split}
    \widehat{\mathbb{P}}(Y^{(a)}=1) &\approx \frac{1}{N} \sum_{j=1}^N \textrm{expit}(\beta_0 + \beta_1 a + \beta_2 c_j)  \\
    \text{where } c_j &\sim \mathscr{N}(\mu_c,\sigma^2_c).
\end{split}
\end{equation}

This can again be used to compute $\Psi_{OR}$ by plugging in $\widehat{\mathbb{P}}(Y^{(1)}=1)$ and $\widehat{\mathbb{P}}(Y^{(0)}=1)$ to \eqref{eq:margor}. As with potential outcome simulation, this does not require specific knowledge of the distribution fo $C$, as long as samples can be generated from it. Note, the estimates of $\mathbb{P}(Y^{(a)}=1)$ could also be used for a risk difference or risk ratio.

We omit simple outcome simulation (with one outcome per row) as a comparator. Were we to use this approach with a dataset of $N$ observations, it would have Monte Carlo error at best as small as potential outcome simulation with $N/2$ observations (where there are two treatment levels). The inequality is clear because, even if the distribution of $C$ were identical under both values of treatment, and the same number of outcomes were simulated under both approaches, potential-outcome simulation conditions on identical $c_i$ under both treatment values, while simple outcome simulation cannot.

\begin{figure}
    \centering
    \includegraphics[width=0.75\linewidth]{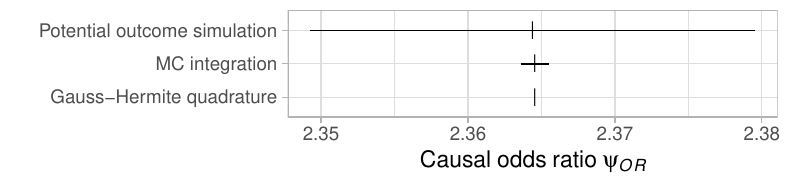}
    \caption{\textbf{Estimates with 95\% prediction intervals for the causal odds ratio with a single confounder}. The plot compares the average estimate using Monte Carlo integration on simulated datasets (labelled ``Potential outcome simulation'') and Monte Carlo integration on the expected value of the outcome (``MC integration'') with the Gauss--Hermite quadrature estimate.}
    \label{fig:ex_1_1d}
\end{figure}

We plot the estimates of the true marginal odds ratio $\Psi_{OR}$ obtained by Gauss--Hermite quadrature and by Monte Carlo integration approaches. The Gauss--Hermite estimate is based on $K=20$ quadrature points. The Monte Carlo integration estimates are based on a simulated dataset of size $N=10^6$. Since the simulation-based methods involve simulating $C$ it is subject to Monte Carlo error, so we repeat the procedure to study its variance. Therefore, $n_{rep}=10^4$ different Monte Carlo estimates are computed, and their corresponding prediction interval is shown in figure~\ref{fig:ex_1_1d}. The estimate computed \textit{via} Gauss--Hermite quadrature is also shown, along with the mean of the Monte Carlo estimates. Gauss--Hermite quadrature utilizes the known Gaussian distribution of $C$; therefore, no simulation is required, which avoids any Monte Carlo error.  The MC estimates based on simulating values of the outcome have a very large variance compared to the MC estimates based on the expected value. The Gauss--Hermite quadrature estimate is instead deterministic, and matches the mean MC estimate very closely. The absolute difference between the Gauss--Hermite estimate and the mean Monte Carlo estimate is $9.51\text{e\,--\,}6$, corresponding to a relative difference of approximately $0.0004\%$. 

\subsection{Multivariate normal confounder}
\label{sec:ex12}
\begin{figure}
    \centering
    \includegraphics[width=0.7\linewidth]{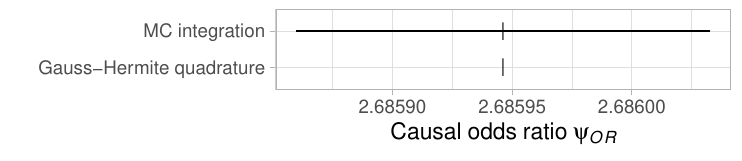}
    \caption{\textbf{Estimates with 95\% prediction intervals for the causal odds ratio with a bivariate confounder}. The plot compares Monte Carlo integration on the expected value of the outcome (``MC integration'') with the Gauss--Hermite quadrature estimate. Potential outcome simulation is omitted because its error is so much higher that it obscures the comparison between the other methods.}
    \label{fig:ex_1_2d}
\end{figure}

Next, we consider a more general distribution for the confounding variable, defined as a vector with two correlated Gaussian variables: 
\begin{align}
\label{eq:2DC}
\begin{bmatrix}
C_1\\
C_2
\end{bmatrix} \,&\sim\, \mathscr{N}\!\left(\begin{bmatrix}
-5\\
-10
\end{bmatrix},\begin{bmatrix}
1 & 1\\
1 & 2
\end{bmatrix}
\right).
\end{align}
The coefficient $\beta_2$ is set to $\bm{\beta}_2 = \begin{bmatrix} 0.1 & 0.1 \end{bmatrix}$. Due to the correlation between the two confounders $C_1$ and $C_2$, we apply the spectral decomposition to transform the grid of quadrature points following the procedure outlined in section \ref{sec:mutiquad}.
As in the previous example, we compare the result of Gauss--Hermite quadrature with the Monte Carlo integration estimate. This is obtained by generating a large number of samples from the distribution of $C_1$ and $C_2$, and taking the sample mean according to equation (\ref{eq:mcex1}) (``MC integration''). The results are shown in figure \ref{fig:ex_1_2d}. Note that the true value of $\Psi_{OR}$ is different from the first example due to the non-collapsibility of the odds ratio \citep{daniel2021making}.

The absolute difference between the Gauss--Hermite estimate and the mean Monte Carlo estimate is $7.59\text{e\,--\,}7$, corresponding to a relative difference of less than $0.00003\%$. 


\subsection{Non-Gaussian confounders}
\label{sec:sim3}
As discussed in section \ref{sec:othergauss}, different quadrature rules can be used to compute results beyond the Gaussian case. We consider three different simulation scenarios, each involving a bivariate confounder $\begin{bmatrix} C_1 & C_2 \end{bmatrix}'$ composed of two independent variables:
\begin{enumerate}
    \item Uniform confounders: $C_1 \sim \textrm{U}(-2,2) \,,~  C_2 \sim \textrm{U}(-4,0)$.
    \item Exponential confounders: $C_1 \sim \textrm{Exp}(1) \,,~  C_2 \sim \textrm{Exp}(2)$.
    \item Gamma confounders: $C_1 \sim \textrm{Ga}(1,2) \,,~  C_2 \sim \textrm{Ga}(4,2)$.
\end{enumerate}
In all the above simulated settings, the outcome probability is set to be 
\begin{equation*}
    \mathbb{P}(Y=1\,\vert\,A,C_1,C_2) \,=\, \textrm{expit}(- A + 0.5\,C_1 + 0.5\,C_2).
\end{equation*}
We design these scenarios in order to have expressions for the corresponding causal odds ratios that can be computed exactly. Additional details on the simulation process are provided in Appendix~\ref{app:nongaussian}.

\begin{figure}
\centering
    \includegraphics[width=.8\linewidth]{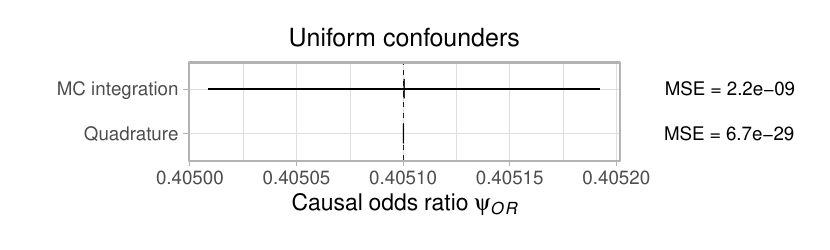}\\
    \includegraphics[width=.8\linewidth]{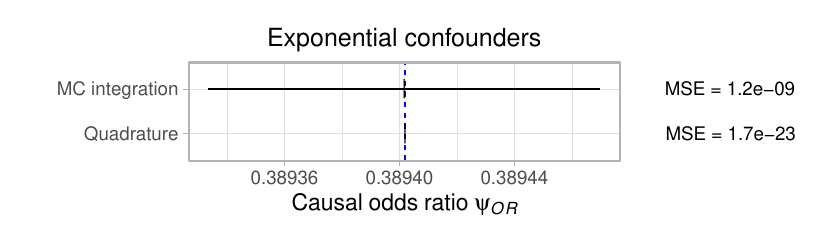}\\
    \includegraphics[width=.8\linewidth]{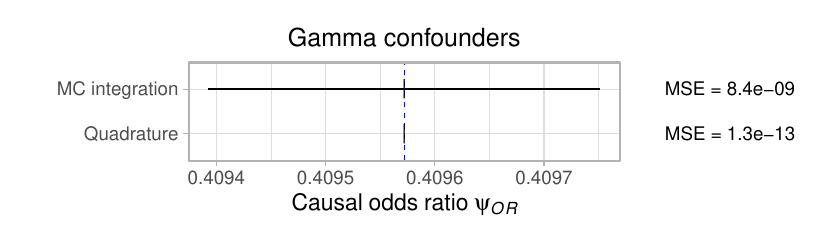}
\caption{Causal odds ratio estimates for two different methods with 95\% prediction intervals for estimating the causal odds ratio under different confounder distributions. The blue dotted line indicates the true causal odds ratio value.}
\label{fig:conf}
\end{figure}

We apply two methods described in section~\ref{sec:confex1}: Monte Carlo integration sampling the outcome probabilities $\mathbb{P}(Y=1\,\vert\,A,C)$ (labelled ``MC integration''), and Gaussian quadrature. All Monte Carlo estimates are based on $N=10^6$ samples; the Gaussian quadrature estimates are based on $K=20$ quadrature points. For each of the three cases, the appropriate quadrature rule is chosen according to table \ref{tab:gaussquads}. The results are shown in figure \ref{fig:conf}, with a line indicating the mean MC estimate based on $10^4$ runs. The MC integration suffers in terms of MSE due to its variance, which is due to the sampling uncertainty of the method. The Gaussian quadrature methods also outperform MC integration in terms of bias: the average of the MC estimates, despite being based on $10^{10}$ samples, exhibit higher bias than the quadrature methods (see also figure \ref{fig:ex1comp}).

\subsection{Convergence and run-time}
\begin{figure}
\centering
    \includegraphics[width=.48\textwidth]{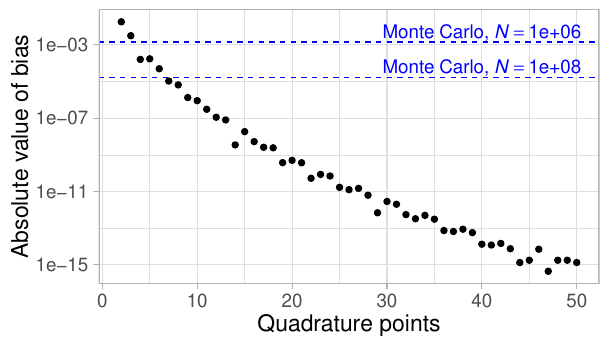}
    \hspace{0.3cm}
    \includegraphics[width=.48\textwidth]{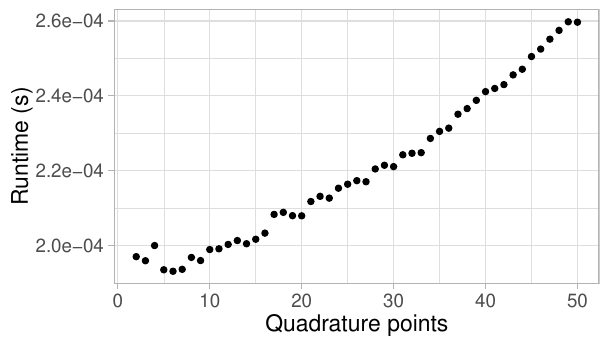}
\caption{\textbf{Bias and run-times of Laguerre quadrature as a function of the increasing number of quadrature points}. The left plot shows the bias (in log scale) as a function of the number of quadrature points. Two Monte Carlo integration estimates are shown in blue as a reference, based on respectively $10^6$ and $10^8$ samples. On the right, the runtime of the Laguerre quadrature averaged over $10^5$ runs. The run-times of the Monte Carlo estimates on the left plot are in the order of 1 and 10 seconds respectively.}
\label{fig:ex1comp}
\end{figure}

Finally, we investigate how the Gaussian quadrature converges to the true value as a function of its quadrature points. As mentioned in section \ref{sec:uniquad}, Gaussian quadrature methods are exact for polynomials $f(x)$ up to order $2K-1$. For smooth functions $f(x)$, Gaussian quadrature features exponential convergence \citep{barrett1961}, while Monte Carlo methods suffer from a constant slow convergence rate $O(N^{-1/2})$. Figure \ref{fig:ex1comp} shows the absolute value of the bias as a function of the increasing number of quadrature points. We use Gauss--Laguerre quadrature due to the closed form expression of the true causal odds ratio in the exponential confounder case. Run-times are averaged over $10^5$ runs. 

The results show that the bias of the Gaussian quadrature indeed decreases exponentially in the number of quadrature points. As a reference, two Monte Carlo integration estimates are shown in blue, based on $10^6$ and $10^8$ samples. The Gaussian quadrature based on 7 quadrature points outperforms the Monte Carlo estimate based on $10^8$ samples. On the right-hand plot, the runtime appears to increase at a near linear rate as a function of the number of quadrature points. Quadratures beyond 50 points are not considered since they approach machine epsilon. The run-times of the MC estimates are not shown as they are in the order of 1 and 10 seconds, being respectively $10^3$ and $10^4$ times slower than Gaussian quadrature. 

Despite the clear advantage of quadrature methods over Monte Carlo in these examples, standard quadrature methods require evaluation at $K$ quadrature points for every dimension to be integrated. These methods therefore suffer from the \textit{curse of dimensionality}, implying that the complexity in evaluating an integral of dimension $D$ will be $\mathcal{O}(K^D)$. Monte Carlo methods can handle high-dimensional integrals with ease, as the complexity of sampling from a multivariate Gaussian distribution is $\mathcal{O}(D^2)$ \citep{thomas08}. Quadrature methods that produce sparse grids able to handle higher dimensions have been proposed \citep{heiss08}, and do not suffer from the curse of dimensionality. These methods are based on sparse constructions that achieve accuracy levels comparable to traditional multivariate grids. 

\begin{figure}
\centering
    \includegraphics[width=.64\linewidth]{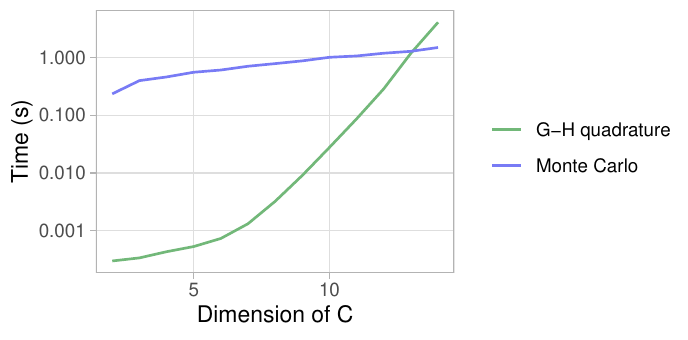}
\caption{Run-times for Gauss--Hermite quadrature \textit{vs.} Monte Carlo integration as a function of the covariate dimension. The number of quadrature grid points $(K=3)$ and MC samples $(N=10^6)$ are chosen to obtain a comparable bias.}
\label{fig:exdim}
\end{figure}

Figure~\ref{fig:exdim} shows the run-times of Gauss--Hermite quadrature and Monte Carlo as a function of the increasing dimension of the confounding variable $C$. The simulation parameters are taken from the example in section \ref{sec:ex12}. As the number of dimensions increases, the runtime of Gauss--Hermite quadrature's increases exponentially, while Monte Carlo increases at a slower rate. 

\section{Applications to mediation}

\subsection{Controlled direct effect}

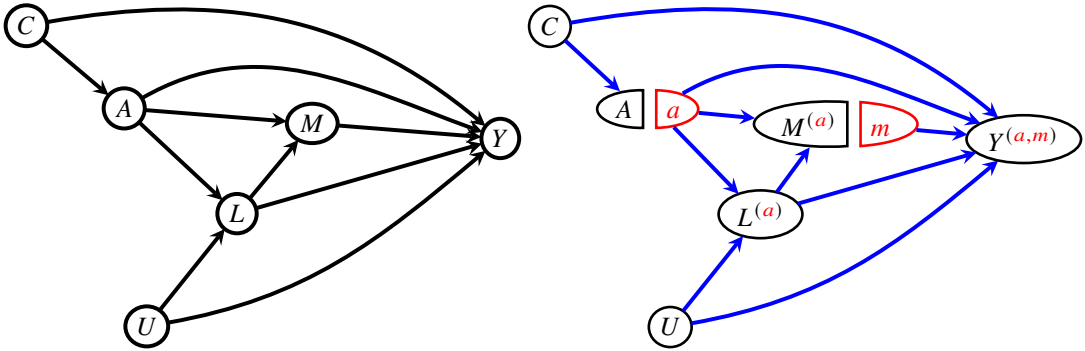
\begin{figure}[htbp]
\centering

\begin{tikzpicture}
\tikzset{line width=1.5pt, outer sep=0pt,
         ell/.style={draw,fill=white, inner sep=2pt,
          line width=1.5pt},
         swig vsplit={gap=5pt,
         inner line width right=0pt}};
\node[name=C, ell, shape=ellipse] at (-.8,1.5) {$C$};
\node[name=A, ell, shape=ellipse] at (.5,.4) {$A$};
\node[name=M, ell, shape=ellipse] at (3,.2) {$M$};
\node[name=L, ell, shape=ellipse] at (2,-1) {$L$};
\node[name=U, ell, shape=ellipse] at (.8,-2.5) {$U$};
\node[name=Y, ell, shape=ellipse] at (5.5,0) {$Y$};

\draw[->,line width=1.5pt,>=stealth,color=black]
(A) edge [out=30,in=160] (Y)
(A) edge (M)
(C) edge (A)
(A) edge (L)
(L) edge (Y)
(L) edge (M)
(U) edge (L)
(U) edge [out=10,in=-140] (Y)
(C) edge [out=10,in=140] (Y)
(M) edge (Y);
\end{tikzpicture}
\begin{tikzpicture}
\tikzset{line width=1pt, outer sep=0pt,
         ell/.style={draw,fill=white, inner sep=2pt,
         line width=1pt},
         swig vsplit={gap=5pt,
         inner line width right=0pt}};

\node[name=A,shape=swig vsplit, swig vsplit={line color right=red}] at (.5,.4) {
    \nodepart{left}{$A$}
    \nodepart{right}{$\textcolor{red}{a}$} 
};
\node[name=M, shape=swig vsplit, swig vsplit={line color right=red}] at (3,.2) {
    \nodepart{left}{$M^{(\textcolor{red}{a})}$}
    \nodepart{right}{$\textcolor{red}{m}$} 
};
\node[name=C, ell, shape=ellipse] at (-.8,1.5) {$C$};
\node[name=L, ell, shape=ellipse] at (2,-1) {$L^{(\textcolor{red}{a})}$};
\node[name=U, ell, shape=ellipse] at (.8,-2.5) {$U$};
\node[name=Y, ell, shape=ellipse] at (5.5,0) {$Y^{(\textcolor{red}{a,m})}$};

\draw[->,line width=1.5pt,>=stealth,color=blue]
(A) edge [out=30,in=160] (Y)
(A) edge (M)
(C) edge (A)
(A) edge (L)
(L) edge (Y)
(L) edge (M)
(U) edge (L)
(U) edge [out=10,in=-140] (Y)
(C) edge [out=10,in=140] (Y)
(M) edge (Y);
\end{tikzpicture}

\caption{Causal diagrams for the controlled direct effect. Left panel: DAG. Right panel: SWIG with interventions $A=a$ and $M=m$\label{fig:cdediag}}
\end{figure}

Next, we revisit the second example of \cite{Naimi2025}. Herein, interest lies in a controlled direct effect, which is the difference in potential outcomes under treatment $a=1$ and control $a=0$ while holding $M$ constant at some value $m$. In a clinical trial, for instance, this could correspond to a hypothetical estimand, which considers what the average treatment effect would be if an intercurrent event $M$ never occurred \citep{iche9}. An example of a controlled direct effect could be the treatment effect had patients never switched from control to experimental treatment during the trial. Such switching is often permitted in oncology clinical trials but would never be part of future care. This estimand can be mathematically defined as:
\begin{align*}
    \Delta_{CDE} =  E[Y^{(a,m)}] - E[Y^{(a^*,m)}],
\end{align*}
where $a\neq a^*$ and we define $Y^{(a,m)}$ as the potential outcome under treatment action $a$ and mediator value $m$. Using the example above, we could be interested in the effect of a treatment in a clinical trial had patients not switched treatment arms (\textit{i.e.} $Y^{(a=1,m=0)}$ \textit{vs.}~$Y^{(a=0,m=0)}$). The structure of the data-generating mechanism for the simulation is depicted by the causal DAG in figure~\ref{fig:cdediag}.

A typical parametric regression model that could be used to generate the outcome  in a simulation is
\begin{align*}
    \Ebb[Y^{(a,m)}|C,A,L,M,U]= g^{-1}(\beta_0+\beta_1a+\beta_2m +\beta_3C + \beta_4L + \beta_5U),
\end{align*}
with some link function $g(\cdot)$. Given this data-generating mechanism (and model), we can express the potential outcome of interest in terms of the observed data using the g-formula \citep{robins1986new}:
\begin{align}
    \Ebb[Y^{(a,m)}] &= \Ebb_C[\Ebb[Y^{(a,m)}|C]] \label{eq:law_iterated_exp}\\
    &= \Ebb_C[\Ebb[Y^{(a,m)}|C,A=a]] \label{eq:exchangeability_a}\\
    &= \Ebb_C[\Ebb_L[\Ebb[Y^{(a,m)}|C,A=a,L]]] \label{eq:iterated_exp_L}\\
    &= \Ebb_C[\Ebb_L[\Ebb[Y^{(a,m)}|C,A=a,L,M=m]]] \label{eq:exchangeability_m}\\
    &= \Ebb_C[\Ebb_L[\Ebb[Y|C,A=a,L,M=m]]] \label{eq:consistency}
\end{align}
The independencies used in the g-formula are $(Y^{(a,m)}\independent A)|C$ invoked in (\ref{eq:exchangeability_a}) and $(Y^{(a,m)}\independent M)|(C,A=a,L)$ invoked in (\ref{eq:exchangeability_m}), which can be read off the template \textit{Single-World Intervention Graph} (SWIG) in figure~\ref{fig:cdediag} \citep{richardson2013single,ocampo2023single}. This describes the worlds we must consider in order to define the controlled direct effect. Otherwise the equalities follow from applying iterated expectation in (\ref{eq:law_iterated_exp}) and (\ref{eq:iterated_exp_L}), as well as the consistency assumption \citep{cole2009consistency} in (\ref{eq:consistency}).

In the simulation we also have $U$, which we can further integrate over to obtain:
\begin{align*}
    \Ebb[Y^{(a,m)}]&=\Ebb_C[\Ebb_L[\Ebb_U[\Ebb[Y|C,A,L,M,U]]]]\\
    &= \Ebb_C[\Ebb_L[\Ebb_U[g^{-1}(\beta_0+\beta_1a+\beta_2m +\beta_3C + \beta_4L + \beta_5U)]]]
\end{align*}

Now suppose the confounders are normally distributed:
\begin{align*}
    C\sim \mathscr{N}(\mu_C,\sigma^2_C),\quad\mu_C &= -10\\
    U\sim \mathscr{N}(\mu_L,\sigma^2_L),\quad\mu_L &= 3 \\
    L|U\sim \mathscr{N}(\mu_U,\sigma^2_U),\quad\mu_U &= 15 + a + 0.1U
\end{align*}

and $\sigma^2_C=\sigma^2_U=\sigma^2_L=1$. Note that $U$ being a cause of $L$ induces correlation between the two. In the actual simulation, we would also need to specify the data-generating mechanisms for $A$ and $M$, but not for the truth calculation because these are set to specific values $a$ and $m$ when considering the specific potential outcomes that define the estimand.

Using the above, we can express the g-formula derived above using the following integrals:
\begin{align}
    \Ebb_C[\Ebb_L[\Ebb_U[\Ebb[Y|C,A,L,M,U]]]] = \int_c\int_\ell\int_u \Ebb[Y|C,A,L,M,U] \phi_U(u)du\phi_L(\ell)d\ell\phi_C(c)dc,
    \label{eqn:fig3}
\end{align}
which we can estimate using multivariate Gauss--Hermite quadrature. 

A consequence of \eqref{eqn:fig3} being a triple integral is that the trivariate Gauss--Hermite quadrature is applied in three-dimensional space. With only $5$ nodes per dimension, the resulting grid includes $5^3$ quadrature points, leading to $125$ summands of the finite sum. 
For this application, it is important that we invoke the spectral decomposition \cite{jackel2005note} to incorporate the correlation between $L$ and $U$. This decomposition rotates the grid according to the known correlation structure, thereby providing more accurate quadrature. 

\begin{figure}
    \centering
    \includegraphics[width=0.95\linewidth]{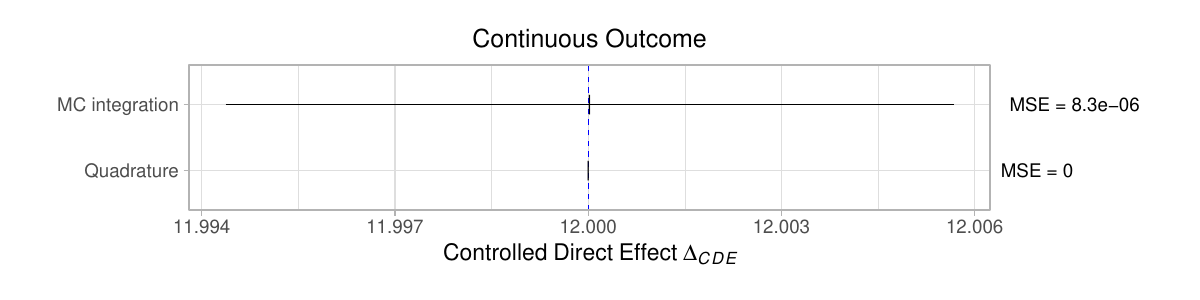}
    \includegraphics[width=0.95\linewidth]{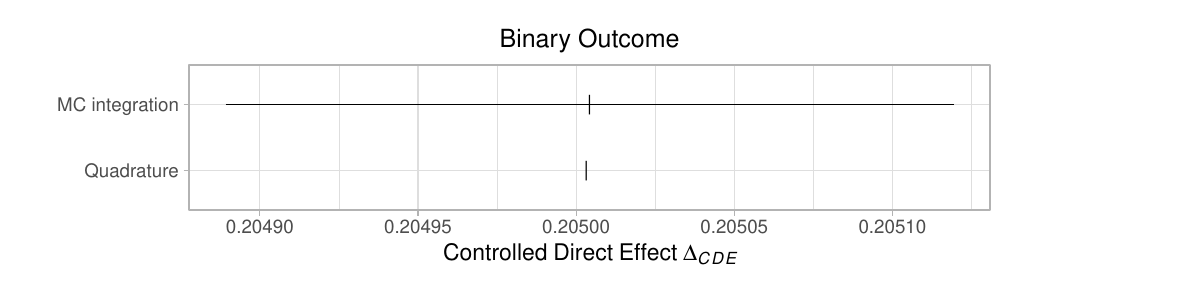}
    \caption{\textbf{Gauss--Hermite Quadrature is more accurate than Monte Carlo Integration in the controlled direct effect simulation.} Result for MC integration shows 95\% prediction interval from $1,000$ repetitions each of size of one million ($10^6$). The mean of the MC integration is the estimated mean based on 1,000 samples of size $10^6$ from the data-generating mechanism averaged across all simulations (i.e. it is equivalent to an MCI with $10^9$ samples). The dashed blue line is the true controlled direct effect in the continuous outcome setting given the data-generating mechanism for the simulation setup.}
    \label{fig:tisim}
\end{figure}

We can see from figure \ref{fig:tisim} that the quadrature again provides an accurate estimate of the truth. We used an identity link for the continuous outcome $Y$, allowing us to discern the true value of $12$, which the quadrature recovers. The MCI also does well, albeit with a bit of Monte Carlo error up to the third decimal point. We also simulated a binary outcome where the truth is unknown. For the binary outcome, the quadrature and MCI are aligned, but the quadrature is deterministic. 


\subsection{Application 3: Mediation for restricted mean survival time}

As presented in \cite{ocampo2024simplifying}, a simulation study was conducted to investigate the properties of using pseudo-values \citep{andersen2010} for mediation with a time-to-event outcome $T$. One of the goals of the simulation study was to investigate bias, MSE, and coverage of the approach for mediation analysis on the restricted mean survival time (RMST) scale. These operating characteristics require knowledge of the true mediation effects. In this paper, we focus on how the true values for the mediation estimands of interest were calculated, which required integrating over the distribution of the mediator.

The simulation setup was as follows, let $A \in \{0, 1\}$ denote the treatment arm, and $M$ the measurement of a continuous mediator. The mediator is drawn from a normal distribution whose mean depends on the value of $a$:
\begin{align*}
    M|(A=a) \sim \mathscr{N}(\mu_{a}, 1),\quad a \in \{0, 1\}.
\end{align*}
For the data generating mechanism, we defined $\mu_{0} = 0$ and $\mu_{1} = -1$. We assume the full effect of the treatment on the mediator happens before any events occur. Then, the time-to-event (TTE) is simulated from an exponential distribution, parameterized as:
\begin{align}
    f(t;\lambda) = \lambda e^{-\lambda t},
    \label{eqn:exp-dens}
\end{align}
where the rate parameter $\lambda$ depends on the treatment and the mediator: 
\begin{align}
    \lambda =  \exp\big(\beta_{0} + A\beta_{A} + M\beta_{M}\big).
    \label{eqn:lambda-model}
\end{align}
Given the exponentially distributed random variable $T$, the true RMST at a given time $\tau$ is:
\begin{align*}
    \mu(\tau; \lambda) = \int_{0}^{\tau} e^{-\lambda t} \diff t = \frac{1}{\lambda}\big(1 - \e^{-\lambda \tau}\big),
\end{align*}

which we can use to define our TE, NDE, and NIE as follows:
\begin{align}
     \text{TE}(\tau) &= \Ebb_{M^{(1)}}\big\{\mu(\tau; a=1, M^{(1)})\big\} - \Ebb_{M^{(0)}}\big\{\mu(\tau; a=0, M^{(0)})\big\},\label{eq:TE}\\
     \text{NDE}(\tau) &= \Ebb_{M^{(0)}}\big\{\mu(\tau; a=1, M^{(0)})\big\} - \Ebb_{M^{(0)}}\big\{\mu(\tau; a=0, M^{(0)})\big\},\label{eq:NDE}\\
     \text{NIE}(\tau) &= \Ebb_{M^{(1)}}\big\{\mu(\tau; a=1, M^{(1)})\big\} - \Ebb_{M^{(0)}}\big\{\mu(\tau; a=1, M^{(0)})\big\},\label{eq:NIE}
\end{align}

For all the above, $M^{(0)} \sim \mathscr{N}(0, 1)$ and $M^{(1)} \sim \mathscr{N}(-1, 1)$. Since all these estimands are expressed in terms of expectations w.r.t a normal random variable (i.e.\@ the mediator under a particular treatment $a$), we can use Gauss--Hermite quadrature detailed in \ref{sec:uniquad} to numerically calculate the true values. These true values of \eqref{eq:TE},\eqref{eq:NDE}, and \eqref{eq:NIE} calculated by quadrature were used to show the approach was unbiased with appropriate coverage probabilities. See section~4 of \cite{ocampo2024simplifying} for more details on the simulation results. While the original simulation focused on a series of scenarios and time points, for the purposes of this paper in comparing the quadrature to MC integration, we look at $\tau=3$. We outline the MCI approach in pseudocode as done by \cite{Naimi2025} for the RMST mediation estimands in \textbf{Algorithm 1}. 
\begin{algorithm}[H]
\caption{Pseudocode for Implementing Monte Carlo Integration for the True RMST Mediation Estimands}
\begin{algorithmic}[1] 
\State set random number generator seed value
\State set large sample size $N$ (e.g. $N=10^9$)
\State simulate $i \in 1 \ldots N$ observations from $M^{(1)}_i\sim\mathscr{N}(\mu_1,\sigma_1)$ and $M^{(0)}_i\sim\mathscr{N}(\mu_0,\sigma_0)$
\State Define $\lambda\big(a,M_i^{(a^*)}\big) = \exp\big(\beta_{0} + a\beta_{A} + M_i^{(a^*)}\beta_{M}\big)$ and compute

$\hat{\mu}_i(\tau, a=1,M_i^{(1)}) = \frac{1}{\lambda(1,M_i^{(1)})}\big(1 - \e^{-\lambda(1,M_i^{(1)}) \tau}\big)$

$\hat{\mu}_i(\tau, a=0,M_i^{(0)}) = \frac{1}{\lambda(0,M_i^{(0)})}\big(1 - \e^{-\lambda(0,M_i^{(0)}) \tau}\big)$

$\hat{\mu}_i(\tau, a=1,M_i^{(0)}) = \frac{1}{\lambda(1,M^{(0)})}\big(1 - \e^{-\lambda(1,M_i^{(0)}) \tau}\big)$

\State Denote the means of these simulated quantities across all subjects $i\in 1 \dots N$ as $\hat{\mu}(\tau,a,M^{(a^*)})=\frac{1}{N} \sum_{i=1}^N \hat{\mu}_i(\tau, a,M_i^{(a^*)})$ and use them to calculate the mediation estimands of interest as:
\begin{align*}
TE(\tau)=  \hat{\mu}(\tau,a=1,M^{(1)}) - \hat{\mu}(\tau,a=0,M^{(0)})\\
NDE(\tau)= \hat{\mu}(\tau,a=1,M^{(0)}) - \hat{\mu}(\tau,a=0,M^{(0)}) \\
NIE(\tau)= \hat{\mu}(\tau,a=1,M^{(1)}) - \hat{\mu}(\tau,a=1,M^{(0)})
\end{align*}
\end{algorithmic}
\end{algorithm}

\begin{figure}
    \centering
    \includegraphics[width=1\linewidth]{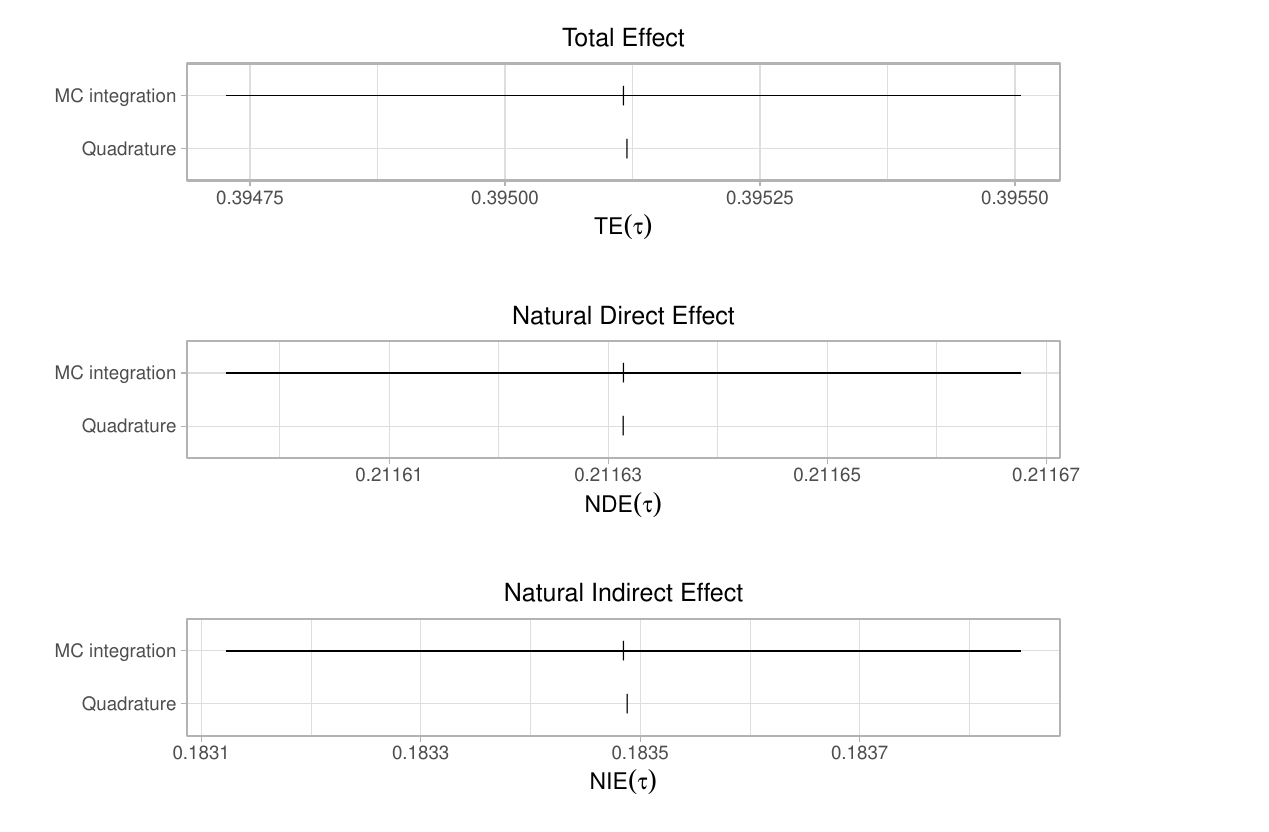}
    \caption{\textbf{Gauss Hermite Quadrature is faster and more accurate than Monte Carlo Integration in computing the true RMST mediation estimands.} Result for MC integration shows 95\% prediction intervals from 1,000 repetitions with dataset size $10^6$. Gauss--Hermite Quadrature is  and aligned with the average of the MCI using $10^9$ samples.}
    \label{fig:rmst}
\end{figure}

When comparing the Gauss-Hermite quadrature to this Monte Carlo integration approach with $10^6$ samples, we again find that the Gaussian quadrature approach is generally more accurate (Figure~\ref{fig:rmst}). The average of the $1000$ repeated MCIs of size $10^6$ converges to the quadrature estimate. Moreover, the MCI took 0.44 seconds on average while the quadrature took 0.073 seconds. While not such a drastic difference for one set of true mediation estimands, given that the original simulation involved 324 ground truths, use of the quadrature led to a more substantial computation time gain in the overarching simulation study than in a single run comparison. 

\subsection{Application 4: Mediation for the hazard ratio}

Consider another mediation simulation study with a time-to-event endpoint, but now instead of the restricted mean survival time, we are interested in performing the mediation analysis on a hazard ratio scale. We begin by again assuming a randomized binary treatment split evenly among the subjects and a mediator comes from a normal distribution $M^{(a)}\sim \mathscr{N}(\mu_a,1)$ where $\mu_a= \alpha_0 - \alpha_aa$. With this, we can draw our time-to-event outcome from a Weibull distribution:
\begin{align*}
    T \sim ~& \text{Weibull}(\gamma,\lambda^\prime), \\
    \text{where } \lambda^\prime = ~& \lambda \exp(\beta_aa+\beta_mm)^\gamma.
\end{align*}
With some algebra, the survival function for this Weibull parameterization can be written as:
\begin{align*}
    S(t) = \exp\left\{- \left(\frac{t}{\lambda}\right)^\gamma \exp(\beta_aa+\beta_mm)\right\}
\end{align*}
and the hazard function as: 
\begin{align*}
    \lambda(t)=\left(\frac{\gamma}{\lambda}\right) \left(\frac{t}{\lambda}\right)^{\gamma -1} \exp(\beta_aa+\beta_mm)
\end{align*}
Using the fundamental relationship $f(t)=\lambda(t)S(t)$, we can write the pdf of $T$ as:
\begin{align*}
    f(t) = \left(\frac{\gamma}{\lambda}\right) \left(\frac{t}{\lambda}\right)^{\gamma -1} \exp(\beta_aa+\beta_mm)\exp\left\{ -\left(\frac{t}{\lambda}\right)^\gamma \exp(\beta_aa+\beta_mm)\right\}
    \label{eq:ltSt}
\end{align*}

In this particular mediation simulation setup, interest lies in the natural direct effect (NDE), natural indirect effect (NIE), and total effect (TE) defined on the hazard ratio scale:
\begin{align}
    NDE = \frac{\lambda(T(1,M^{(0)})}{\lambda(T(0,M^{(0)})}\\
    NIE = \frac{\lambda(T(1,M^{(1)})}{\lambda(T(1,M^{(0)})} \\
    TE = \frac{\lambda(T(1,M^{(1)})}{\lambda(T(0,M^{(0)})}
    \label{eq:NEs}
\end{align}
None of these terms correspond to any of the parameters (or combinations thereof) used to parameterize the data-generating mechanism for the simulation study. Since the hazard is not expressible as an expectation, we cannot calculate the marginal value of the hazard by integrating the $\lambda(t;a,m)$ over the distribution of the mediator. Instead, we use the the fact that $ \lambda(t)=\frac{f(t)}{S(t)}$ to compute the true values of these estimands. We can identify the numerator $f(t)$ using the non-parametric mediation formula \citep{imai2010general},
\begin{align}
    f(T_i(a,M^{(a^\prime)})=t) &= \int_{\mathscr{M}(a^\prime)} f(t_i|A_i=a,M_i=m)dF_{M^{(a^\prime)}}(m|A_i=a^\prime) \\
    &=  \int_{\mathscr{M}(a^\prime)} f(t_i|A_i=a,M_i=m) \phi_{M^{(a^\prime)}}(m|A_i=a^\prime)dm,
    \label{eq:ft}
\end{align}
which can be applied to derive the pdf of any of the four possible nested counterfactuals $T(1,M(1))$, $T(0,M(0))$, $T(1,M(0))$, or $T(0,M(1))$ used to define the mediation estimands above. Due to knowledge of $f(t)$ defined in~\eqref{eq:ltSt}, and to the normality of $M(a)$, this can be computed \textit{via} Gauss--Hermite Quadrature. Similarly, the denominator $S(t)$ can be written as:
\begin{align}
    S(T_i(a,M^{(a^\prime)})=t) = \int_{\mathscr{M(a^\prime)}} S(t|A_i=a,M_i=m)\phi(m|A_i=a^\prime)dm
    \label{eq:St}
\end{align}
by iterative expectation. This can also be solved with Gauss--Hermite quadrature. Dividing \eqref{eq:ft} by \eqref{eq:St} gives us the components required to compute our mediation estimands \eqref{eq:NEs}. By calculating these functionals across the range of plausible values of $t$ and averaging, we can obtain the desired mediation effects.

\section{Discussion}

Simulation studies are such a powerful widely-used tool in statistical research because the researcher knows, and has some control over, the data-generating mechanism. Historically, simulation studies tended to focus on ``parameter recovery'': estimands of interest were conveniently defined as parameters of the data-generating mechanism. Increasing appreciation that estimands should be defined nonparametrically -- that is, outside of a statistical model -- means parameter recovery is becoming a less ubiquitous aim~\citep{iche9}. In developing a simulation study, the development of data-generating mechanisms frequently involves range-finding, often done by trial-and-error. Armed with a fast way to compute the true value of an estimand (e.g. Gaussian quadrature), researchers can more efficiently pick their desired simulation parameters. That the true value is accurate matters for simulation performance measures that use the truth in their definitions \citep[see][table~6]{Morris2019}. If the true value is subject to small errors, this impacts estimated performance, giving the appearance of bias, higher mean squared error, undercoverage, and so on.

The estimand is not always expressible in terms of parameters of the data-generating mechanism. In such cases, we have discussed the following four options:
\begin{description}
    \item[\textbf{Outcome simulation}]conceptually straightforward but relatively slow even in simple cases, and has high Monte Carlo error \citep{Morris2019};
    \item[\textbf{Potential outcome simulation}]similar to outcome simulation, but reduces Monte Carlo error by simulating both (or all) potential outcomes per row.
    \item[\textbf{Monte Carlo integration}]similar to potential outcome simulation but further reduces Monte Carlo error by calculating expected values instead of drawing realisations; again conceptually straightforward and relatively slow \citep{Naimi2025}, and works when the estimand is a function of potential outcome means;
    \item[\textbf{Quadrature}]No Monte Carlo error and instantaneous in simple cases but can become slow with many dimensions; accurate provided sufficient quadrature points $K$ and that the correlation structure is taken into account \textit{via} Cholesky or Spectral decomposition.
\end{description}
Despite these attractive features, we acknowledge some disadvantages. First, in higher dimensions, quadrature retains its accuracy but becomes slower than Monte Carlo integration. Second, quadrature will not always be applicable. For example, because it utilizes special knowledge of the covariate distributions, it could not be used for so-called ``plasmode'' simulation, which uses observed covariates of unknown distribution \citep{Schreck2024,shaw2025}. Third, quadrature involves some algebra to obtain the functional and determine the integrand, which is not required for outcome simulation or Monte Carlo integration. Researchers may lack confidence to perform these calculations as opposed to simulation-based approaches. Spending time employing quadratures may be a worthwhile effort when considering many data-generating mechanisms, as the quadrature can provide a huge time-saving. Additionally, due to their deterministic nature,  quadrature methods are extremely memory-efficient. Precise MC estimates on the other hand rely on storing large sets of sampled data, leading to a much higher memory usage, which can quickly become an issue especially when the underlying data-generating mechanism carries a high variance. 

One practical question is where to report this in a simulation study. It could reasonably be under ``Data-generating mechanism'' (it is the true value implied by that mechanism), ``estimand'' (it is the true value of an estimand) or even ``performance measures'' (the true value is required to be able to compute measures such as bias and coverage). Our view is that ``estimand'' is most appropriate, but that details should ultimately appear in an appendix or supplementary materials; calculation of the truth needs to be correct, but is a nuisance that could detract from other aspects of of the simulation study at-hand.

\bigskip \noindent {\bf{Acknowledgement}}

\noindent {\it Many thanks to Novartis colleagues in the Neuroscience knowledge sharing meeting for feedback and discussions.}

\bigskip \noindent {\bf{Conflict of Interest}}

\noindent {\it{TPM  owns shares in Novartis Pharma AG. He has received consultancy fees from Novartis Pharma AG, Bayer Healthcare Pharmaceuticals, Alliance Pharmaceuticals, Gilead Sciences, Kite Pharma, and has received income for teaching about simulation studies}}

%% file: appendices.tex
\newpage
\appendix

\newpage
\section{Quadrature rules for non-Gaussian distributions \label{app:nongaussian}}
This section explores different types of quadrature beyond the Gauss--Hermite case to estimate expectations of the type $\int\!f(x)\,p(x)\,dx$, where $p(x)$ is the density of a (possibly multivariate) variable with respect to $f(x)$ is integrated. Table \ref{tab:gaussquads} in the main text introduces three types of quadrature, each corresponding to a different density $p(x)$. We provide details for the implementation of each of these quadrature rules to estimate causal quantities for simulation studies.

\subsection*{Uniformly distributed variables}
In simulation studies, it may be that the measured/unmeasured confounders that we need to integrate over have a uniform distribution.
In the case of a uniform density $p(x)\propto 1$ on the interval $[-1,1]$, the optimal Gaussian quadrature points $x_i$ are given by the roots of the Legendre polynomials $P_n(x)$. In accordance to the Gaussian quadrature construction, the weights are computed in order to provide exact results for polynomials $f(x)$ up to order $2n-1$. The associated weights $w_i$ are 
\begin{equation*}
    w_i \,=\, \frac{2}{(1-x_i^2)\left[P'_n(x_i)\right]^2}\,.
\end{equation*}
The quadrature points are associated weights are designed for integrals of the type $\int_{-1}^1f(x)dx$. For generic uniform densities on the interval $[a,b]$, the integral can be rewritten \textit{via} the simple change of variable $t=\frac{2x-b-a}{b-a}$:
\begin{align*}
    \int_a^b f(x)\,dx \,=\, 
    \frac{b-a}{2}\int_{-1}^{1} f\!\left(\frac{b-a}{2}t + \frac{b+a}{2}\right) dt.
\end{align*}
Computing the causal odds ratio in the simulation settings described in section \ref{sec:sim3} requires computing the following integrals for $a=\{0,1\}$:
\begin{equation*}
    \mathbb{P}(Y^{(a)}=1) = \int_0^\infty\!\!\int_0^\infty \frac{1}{16} \,\textrm{expit}\left(\frac{c_1}{2}+\frac{c_2}{2}-a\right) dc_1 dc_2.
\end{equation*}
The exact results can be written in terms of the dilogarithm function $\textrm{Li}_2(z)$, which can be computed with arbitrary precision:
\begin{align*}
    \mathbb{P}(Y^{(0)}=1) \,&=\, \frac{1}{4}\left[2\,\textrm{Li}_2(-e^{-1})-\textrm{Li}_2(-e)-\textrm{Li}_2(-e^{-3})\right] \quad\textrm{and} \\
     \mathbb{P}(Y^{(1)}=1) \,&=\,\frac{1}{8}\left[4\,\textrm{Li}_2(-e^{-2})-2\,\textrm{Li}_2(-e^{-4})+\frac{\pi^2}{6}\right]\!.
\end{align*}

\subsection*{Exponentially distributed variables}
In the case of an exponential density $p(x)=e^{-x}$, the optimal Gaussian quadrature is given by the Laguerre polynomials $L_n(x)$. In particular, each quadrature point $x_i$ is given by the $i$-th root of the Laguerre polynomial $L_n(x)$, and the corresponding weights $w_i$ are 
\begin{equation*}
    w_i \,=\, \frac{1}{x_i\left[L'_n(x_i)\right]^2}\,.
\end{equation*}
The quadrature points are associated weights are designed for integrals of the type $\int_0^\infty e^{-x} f(x)dx$. For generic exponential densities $p(x) = \lambda e^{-\lambda x}$, the integral can be rewritten \textit{via} the simple change of variable $t=\lambda x$:
\begin{equation*}
    \int_0^\infty f(x)\,\lambda e^{-\lambda x}\, dx \,=\, \int_0^\infty f\!\left(\frac{t}{\lambda}\right) e^{-t} \,dt.
\end{equation*}
Computing the causal odds ratio in the simulation settings described in section \ref{sec:sim3} requires computing the following integrals for $a=\{0,1\}$:
\begin{equation*}
    \mathbb{P}(Y^{(a)}=1) = \int_0^\infty\!\!\int_0^\infty 2e^{-2c_2}\,e^{-c_1}\,\textrm{expit}\left(\frac{c_1}{2}+\frac{c_2}{2}-a\right) dc_1 dc_2.
\end{equation*}
The above integrals have a closed form expression, with the exact results being
\begin{align*}
    \mathbb{P}(Y^{(0)}=1) \,=\, \frac{2}{3}~~\quad\textrm{and}~~\quad
     \mathbb{P}(Y^{(1)}=1) \,=\, \frac{4}{e}\left[\frac{2}{3}+\frac{1}{e}\left(\frac{1}{2}-\frac{1}{e}+\frac{1-e^2}{e^2}\log(1+e)\right)\right]\!.
\end{align*}

\subsection*{Gamma-distributed variables}
A more general quadrature rule holds for densities with a Gamma kernel $p(x) \propto x^\alpha e^{-x}$, \textit{via} the generalized Laguerre polynomials $L_n^{(\alpha)}(x)$ \citep{Rabinowitz59}. The corresponding weights $w_i$ are 
\begin{equation*}
    w_i \,=\, \frac{x_i\,\Gamma(n+\alpha+1)}{n!\,(n+1)^2\,[L_{n+1}^{(\alpha)}(x_i)]^2}\,.
\end{equation*}
The quadrature points are associated weights are designed for integrals of the type $\int_0^\infty x^\alpha e^{-x} f(x)dx$. For general gamma densities $p(x)\propto x^\alpha e^{-\lambda x}$, the integral can be rewritten by \textit{via} the same change of variable $t=\lambda x$:
\begin{equation*}
    \int_0^\infty f(x)\,\frac{\lambda^{\alpha+1}}{\Gamma(\alpha+1)} x^\alpha e^{-\lambda x}\, dx \,=\, 
    \frac{1}{\Gamma(\alpha+1)} \int_0^\infty f\!\left(\frac{t}{\lambda}\right)t^\alpha\,e^{-t}  \,dt.
\end{equation*}

Computing the causal odds ratio in the simulation settings described in section \ref{sec:sim3} with $C_1 \sim \textrm{Ga}(1,2) ~\textrm{and}~ C_2 \sim \textrm{Ga}(4,2)$ requires computing the following integrals for $a=\{0,1\}$:
\begin{equation*}
    \mathbb{P}(Y^{(a)}=1) = \int_0^\infty\!\!\int_0^\infty \frac{2^5}{\Gamma(4)}\, e^{-2c_1}\,c_2^3\,e^{-2c_2} \,\textrm{expit}\left(\frac{c_1}{2}+\frac{c_2}{2}-a\right) dc_1 dc_2.
\end{equation*}
To obtain a closed form expression, the integral can be simplified using the fact that $Z=0.5\,C_1+0.5\,C_2 \sim \textrm{Ga}(5,1)$:
\begin{equation*}
    \mathbb{P}(Y^{(a)}=1) = \int_0^\infty  \frac{1}{\Gamma(5)}\, z^4\,e^{-z} \,\textrm{expit}\left(z-a\right)\,dz.
\end{equation*}
The exact results are then
\begin{align*}
    \mathbb{P}(Y^{(0)}=1) \,=\, \frac{15}{16}\,\zeta(5)~~\quad\textrm{and}~~\quad
     \mathbb{P}(Y^{(1)}=1) \,=\, \frac{3+10\,\pi^2+7\,\pi^4-360\,\textrm{Li}_5\!\left(-e^{-1}\right)}{360\,e}
\end{align*}
where $\textrm{Li}_5(z)$ is the base 5 polylogarithm function, and $\zeta$ is the Riemann zeta function, both of which can be computed with arbitrary precision.

Multivariate extensions of the above results are straightforward, with the \texttt{mvQuad} R package \citep{citemvquad} offering implementations for multivariate Gauss--Legendre and Gauss--Laguerre quadrature rules. For the generalized Gauss--Laguerre quadrature, the \texttt{gaussquad} R package \citep{citegaussquad} provides functions to calculate the weights and quadrature points.